\documentclass{article}

\usepackage{arxiv}

\usepackage[utf8]{inputenc} 
\usepackage[T1]{fontenc}    
\usepackage{hyperref}       
\usepackage{url}            
\usepackage{booktabs}       
\usepackage{amsfonts}       
\usepackage{nicefrac}       
\usepackage{microtype}      
\usepackage{lipsum}
\usepackage{graphicx}
\usepackage{multirow}
\usepackage{lscape}
\usepackage{xcolor}
\usepackage{colortbl}
\usepackage{caption}
\usepackage{subcaption}

\title{A Two-Week In-the-Wild Study of Screen Filters and Camera Sliders for Smartphone Privacy in Public Spaces}

\author{Andreas Tjeldflaat* \\
    University of Bergen \\
    Bergen, Norway\\
    \texttt{a.tjeldflaat@gmail.com}
\And
 Piero Romare* \\
  Chalmers University of Technology \\
  University of Gothenburg \\
  Gothenburg, Sweden \\
  \texttt{pieror@chalmers.se} \\
\And
 Yuki Onishi \\
  University of Bergen\\
  Bergen, Norway \\
  \texttt{yuki.onishi@uib.no} \\
\And
  Morten Fjeld \\
  t2i Lab, CSE, Chalmers University of Technology \\
  Gothenburg, Sweden \\
  \texttt{fjeld@chalmers.se} \\
  \And
  Bjørn Sætrevik \\
  University of Bergen \\
  Bergen, Norway \\
  \texttt{bjorn.satrevik@uib.no}
}

\begin{document}
\maketitle
\begin{abstract}
Smartphone usage in public spaces can raise privacy concerns, in terms of shoulder surfing and unintended camera capture. In real-world public space settings, we investigated the impact of tangible privacy-enhancing tools (here: screen filter and camera slider) on smartphone users' reported privacy perception, behavioral adaptations, usability and social dynamics. We conducted a mixed-method, in-the-wild study ($N = 22$) using off-the-shelf smartphone privacy tools. We investigated subjective behavioral transition by combining questionnaires with semi-structured interviews. Participants used the screen filter and the camera slider for two weeks; they reported changes in attitude and behavior after using a screen filter including screen visibility and comfort when using phones publicly. They explained decreased privacy-protective behaviors, such as actively covering their screens, suggesting a shift in perceived risk. Qualitative findings about the camera slider suggested underlying psychological mechanisms, including privacy awareness and concerns about social perception, while also offering insights regarding the tools' effectiveness.
\end{abstract}


\section{Introduction}\label{sec:introduction} 
\begin{figure*}[!ht]
    \centering
    \includegraphics[width=\linewidth]{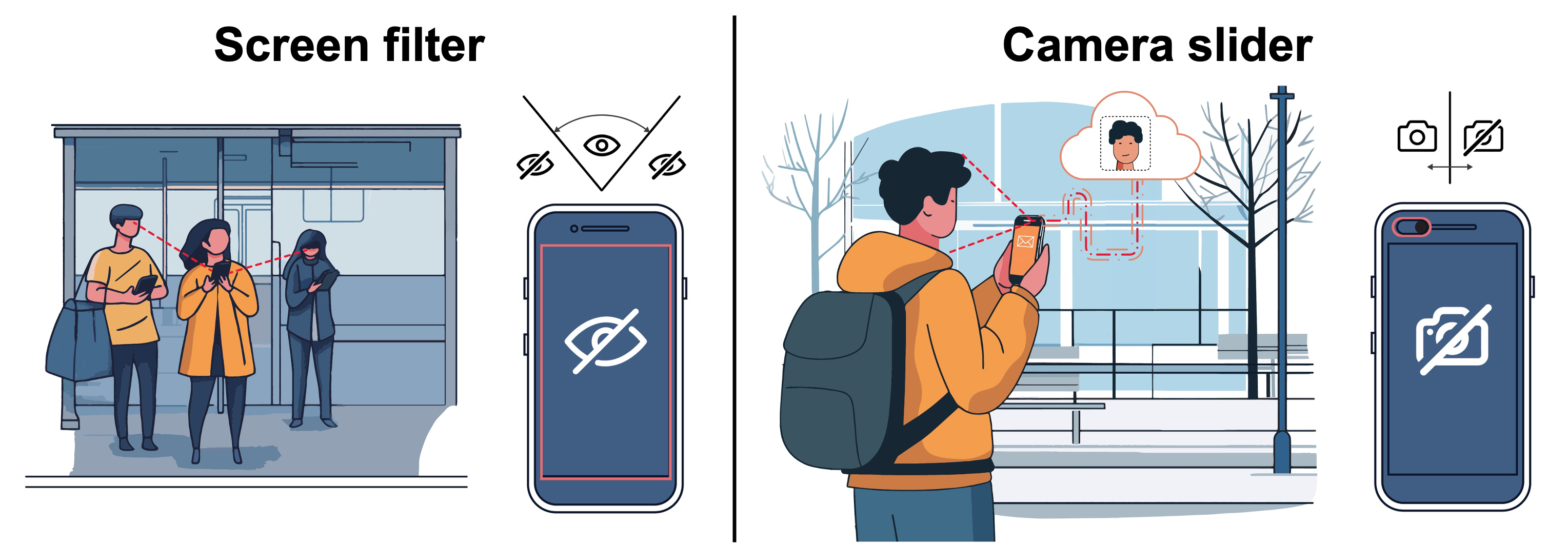}
    \caption{Everyday privacy challenges smartphone users face in public spaces, from physical intrusions like shoulder surfing (left) to invisible risks like unauthorized camera access and online data sharing (right). In addressing such challenges, we studied two tangible privacy tools; privacy screen filter (left) and camera slider (right).}
    \label{fig:placeholder}
\end{figure*}

In a digital world, privacy is no longer an occasional or abstract concept, but a constant, everyday challenge when interacting with smartphones in public. From smartphones to public Wi-Fi, individuals are usually around people and exposed to online tracking systems leading to privacy threats~\cite{li2016csi, wang2023accurately, ali2019privacy, cunche2014know}. This double exposure complicates how individuals assess and manage their privacy risks, particularly in overlapping physical and digital contexts where boundaries of information sharing is not always clear. At the same time, when people access services from their smartphone, their primary goal is usually to get quick access to the content or functionality rather than to manage their privacy. Indeed, privacy is often seen as a secondary concern~\cite{fischer2024curious}.

There is an increasing demand to explore more intuitive and tangible forms of interaction that treat privacy not merely as a configurable setting but as a clear, actionable component of the primary user interface, motivating users to actively engage with tangible privacy mechanisms supported by direct feedback~\cite{delgado2024you, mehta2019tangible}. Tangible and embodied solutions may support intuitive privacy management by integrating protective functions directly into everyday interactions. The adoption and utilization of online connected devices is often characterized by attitudes such as privacy concerns~\cite{dienlin2015privacy}, perceived benefits and risks~\cite{koohikamali2015location}, trust~\cite{zhang2014effects} and perceived control~\cite{xu2012research}, behavior including protection behavior and information disclosure behavior~\cite{gerber2018explaining} and social norms driven also by cultural influences~\cite{beldad2015here,van2015share} shaping privacy protection strategies. These strategies are shaped not only by personal beliefs but also by contextual factors, including the physical environment and perceived social acceptability. Thus, we frame our analysis using the Contextual Integrity (CI) framework, which defines privacy as the appropriate flow of information governed by specific transmission principles and attributes relevant to the context of smartphone use in public spaces.
The camera slider or cover (see Figures~\ref{fig:placeholder} and~\ref{fig:tools}, right) protects individuals from unauthorized access to their face images. Attitudes, social environment, and
perceived control over protecting privacy influence the users behaviour and decision over laptop webcam covering as shown with questionnaires~\cite{machuletz2018webcam,machuletz2016users}. 
The privacy screen filter or film (see Figures~\ref{fig:placeholder} and~\ref{fig:tools}, left) can be used in everyday life for protecting privacy from shoulder surfing, and in specific contexts, such as schools, to avoid cheating~\cite{escudier2014student}. By increasing the spatial frequency and decreasing the luminance contrast of visualization in a screen, it is harder for the onlooker to clearly see the details in a smartphone from a tested distance of $\approx 60-90$ cm, while remaining unchanged when $\approx 30$ cm, showing the effectiveness of privacy protection and its usability~\cite{zhang2023don,chen2019keep}. Similarly, to protect smartphone photos from onlookers, different graphical obfuscation methods were tested with the acceptability evaluated through correct identification of the obfuscated photos by the owners~\cite{von2016you}. The Eye-shield was a software solution designed to protect images and texts in that regard, reducing recognition rates of 24\% and 16\%~\cite{tang2023eye}. 
However, a recent SoK investigated the existing protection mechanisms against shoulder surfing, highlighting users' preference for non-digital alternatives~\cite{farzand2024sok}. 

In this paper, we explore the real-world impact of tangible smartphone privacy tools by focusing on two widely accessible solutions. The paper is motivated by the idea that tangible privacy tools, privacy screen filters and camera sliders, are different because of their physicality and their mechanisms influencing privacy management beyond software solutions. We argue that this physicality introduces dimensions such as control, awareness, and social interpretation, transforming how users negotiate privacy in public spaces. These tools also differ in how users engage with them: the screen filter represents passive tangible privacy, offering protection from the surrounding presence without requiring ongoing user action, in contrast to the camera slider which reflects active tangible privacy protection from online surveillance, demanding deliberate user intervention (see Figure~\ref{fig:placeholder}). We investigate users’ perceptions and everyday practices surrounding these tools, framing our discussion under the lens of CI~\cite{nissenbaum2004privacy}.

\paragraph{Contributions} We preliminarily explored the impact of tangible privacy tools on engaging privacy perception through a mixed-method in-the-wild study with two types of off-the-shelf privacy tools for mobile devices using CI. The results show the following:
\begin{itemize}
    \item The impact of privacy mobile device screen filters on enhancing the privacy perception and behavioral adaptation in public physical environments.
    \item The insights of social and usability challenges of the mobile device front camera slider in in public and/or social settings. 
    \item The design interpretations on how physical usability, social perception, and contextual awareness can be balanced to support effective privacy protection.
\end{itemize}
\section{Background and Related Work}\label{sec:related_work}
This section presents a background  the Contextual Integrity (CI) framework and reviews relevant existing literature.
It explores the emerging field of tangible privacy tools, before specifically examining existing physical privacy protections for smartphones, such as screen filters and camera sliders, which are the focus of this paper's investigation.

\subsection{Contextual Integrity}\label{sec:contextual_integrity}
CI was elaborated by Nissenbaum~\cite{nissenbaum2004privacy}, who defines privacy as contextual information-sharing expectation. CI identifies five parameters that are involved in structuring these norms: 1) data subject, 2) sender, 3) receiver, 4) attribute (e.g., type of information) and 5) transmission principle (e.g., conditions where the information is shared). These parameters have been widely applied to study usable privacy~\cite{malkin2022contextual}. They are applied to understand the privacy implications from a sociotechnical perspective~\cite{wisniewski2022privacy,benthall2017contextual}. In several contexts, studies related to information flow has been evaluated with regard to societal and contextual privacy norms to extract privacy violations when such an information flow differs from the social norms~\cite{kumar2024roadmap}. These include wearables sharing data where weight and location was perceived as less appropriate than sharing health-related attributes like heart disease risk~\cite{bourgeus2024understanding}; smart voice assistance~\cite{brause2024there}; Virtual Reality (VR) classrooms that introduce new information flow within the classroom context to identify potential violations of established privacy norms~\cite{brehm2023understanding}; virtual classrooms~\cite{cohney2021virtual}; and video conferences where the privacy concerns include the features related to user-tracking, such as attention, pinning and spotlighting~\cite{kim2025understanding}. These studies highlight the principle that, within different contexts and technologies, users’ expectations of appropriate information flow depend on the sensitivity of the attribute and the context of disclosure. In this study that employed tangible privacy tools, we adopt CI as an established framework to facilitate the operationalization of privacy into concrete and observable events~\cite{benthall2017contextual,barkhuus2012mismeasurement}.

\subsection{Tangible and Embodied Privacy Tools}\label{sec:tangible_embodied}
Research on tangible privacy is emerging and gaining scholarly attention. For example a data(pay)slip could promote user awareness about their data sharing to enhance transparency~\cite{gomez2024dataslip}. The Privacy Care Framework~\cite{mehta2021privacy} fostered awareness and control through familiar bodily gestures, engaging haptic senses to ease privacy management. Further, ~\cite{ahmad2022tangible} studied smart voice assistants and found that devices with physical privacy controls were seen as more trustworthy and usable than those with only non-tangible mechanisms. Indeed users would prefer “need for touch” and technological affinity~\cite{delgado2024you}. Rodriguez et al.~\cite{rodriguez2021take} noted that many privacy assistants lack intuitiveness and engagement, especially for less technical users, supporting tangible elements to improve usability. PriKey~\cite{delgado2022experiencing}, a device-independent tangible privacy mechanism for smart homes, provides sensor-based, user-centric control over privacy invasive devices. Ahmad et al.~\cite{ahmad2020tangible} examined how bystanders would manage privacy risks in IoT applying CI and the Altman boundary regulation theory, highlighting reliance on physical environments and the need for device transparency~\cite{altman1975environment}.

Since webcam status LED are insufficient to deliver the recording ongoing~\cite{portnoff2015somebody}, Eyecam proposes an aesthetic artifact that may increase individuals’ awareness of surveillance, with multiple roles including mediator, observer, mirror, presence, and agent~\cite{teyssier2021eyecam}. Webcam control and hardware toggles have been studied for smart voice assistant and smart home~\cite{ahmad2022tangible,shalawadi2024manual} without considering the novelty effect. Regarding shoulder surfing, a few studies considered the longitudinal design to analyzed protection mechanisms~\cite{farzand2025understanding} and authentication related threats~\cite{harbach2014s}. A diary study has used as intervention design to understand who is a shoulder surfer, where and when shoulder surfing can happen~\cite{farzand2022shoulder}. Most research relies on questionnaire or interviews~\cite{farzand2025understanding}, with limited focus on behavioral impacts of tangible privacy interventions, especially in dynamic, public mobile device contexts. This paper explores the impact of tangible privacy tools through a two-week intervention in public contexts using questionnaire and interview to assess longitudinal behavioral adaptation or social environment effects.

\subsection{Smartphone Tangible Privacy Tools}\label{sec:privacy_embodied}
Existing physical privacy protections can serve as tangible privacy tools for smartphones, including microphone blockers against jamming attacks~\cite{chen2020wearable}, RFID blockers protecting credit cards from skimming~\cite{spiekermann2009rfid}, two-factor authentication tokens~\cite{lang2016security,krol2015they,reese2019usability,reynolds2018tale}, screen filters, and camera sliders. 

Screen filters help prevent shoulder surfing, a visual privacy invasion where a nearby person views a smartphone "without consent"~\cite{farzand2021interplay}. Shoulder surfing violates the CI transmission principle as observers intercept information without consent. In public displays, the shoulder surfing is created from curiosity rather than bad intentions, since the larger is the display the more honey-pot effect is in place~\cite{brudy2014anyone}. Considering desktop monitors and by deceiving participants, a memory test measured that they were likely to read sensitive text on larger display after finishing a small fake survey~\cite{tan2003information}. A 2016 survey found 31\% of participants admitted to overlooking others’ phones within a year~\cite{marques2016snooping}, and this likely increased with smartphone diffusion. Shoulder surfing risks include identity theft, stalking, and device threats, also impacting others, such as parents and children~\cite{farzand2024you}. A one-month diary study showed shoulder surfing often occurs without the victim's awareness, mostly on public transport and workplaces~\cite{farzand2022shoulder}. Indeed, by proposing public transport scenario the visual attention using eye-tracking was measured resulting that 4 out of 16 participants where able to recall the authentication pattern and that overall all the participants looked at least once at the target phone~\cite{saad2021understanding}. EyeSpot~\cite{khamis2018eyespot} suggested a smartphone screen hiding technique using eye-tracking that show the are of the screen content based on the user's attention. VR scenarios further revealed a behavioral attack model with idle, approach, and attack stages~\cite{abdrabou2022understanding}. Depending on the viewing angle, privacy screen filters can prevent shoulder surfing from accessing sensitive information. While it can happen by accident, thus without explicit purpose rather than being bored~\cite{harbach2014s}, it can also led to observation attacks in the scenario where the device owner is inserting authentication credentials~\cite{eiband2017understanding}. An easy-to-learn and use smartphone PIN system solely based on auditory or tactile stimuli has been proposed to avoid shoulder surfing ~\cite{bianchi2010phone}, and for pairing the phone to external systems while preventing similar observation attacks~\cite{bianchi2010authentication}. Several schemes including spatial, temporal and visual cues have been studied to protect credentials against shoulder surfing attacks~\cite{bovsnjak2020shoulder}. Considering the shoulder surfing perception in a questionnaire and longitudinal field study showing that 65\% of participants of their questionnaire were not particularly concern about getting steal their credentials, and, if in place, others explained five measures against shoulder surfing including tilt their screen, wait the right moment, turn, covert the phone or change their unlock code. Considering the field study, only the $0.3\%$ of the cases, the participants reported a shoulder surfer looking at their phone~\cite{harbach2014s}. Findings protection strategies against shoulder surfing was the aim of~\cite{kuhn2019user}, finding a total of seven strategies along awareness and effort dimensions. Our paper evaluates young adults’ perceived privacy and behavioral adaptations toward screen filters in a two-week in-field study as partial shoulder surfing prevention.

Camera sliders offer a simple, effective means to enhance perceived privacy. Manual camera sliders tend to be trusted more than hybrid or automated options, but their long-term challenges include remembering consistent privacy actions~\cite{do2021smart}. Despite the potential user load reduction of automation, reluctance to put it into practice persists~\cite{shalawadi2024manual}. We contribute by reporting social and usability challenges of manual camera sliders.

The unique challenge of public mobile device use lies in the constant negotiation between usability and privacy in dynamic shared spaces. Users often adapt their behavior, such as angling screens away from others or covering cameras manually, depending on situational context, social norms, and perceived risk. Furthermore, while some studies examine tangible tools like webcam sliders, few intervention-based studies have assessed actual behavioral adaptation over time or in real-world social environments. Our study directly addresses this by conducting a two-week intervention with smartphone users in their everyday public contexts, moving beyond single attitudinal indication to report changes in behavior, applying CI as encouraged by Barkhuus~\cite{barkhuus2012mismeasurement}.

\section{Methods}\label{sec:methods}
We explored how tangible privacy tools for smartphone affect users' perceived privacy and behavioral adaptations in everyday public and social settings as well as the usability challenges. These tools can be broadly characterized by their way of interaction: passive tangible privacy, such as the privacy screen filter, operates without requiring continuous user intervention; while active tangible privacy, the camera slider, demands deliberate user action to engage or not the privacy protection. To explore the impact of tangible privacy tools, we set the following Research Questions (RQs):
\begin{itemize}
    \item[\textbf{RQ1a}]: How does the level of privacy perception change with the privacy screen filter in public spaces?
    \item[\textbf{RQ1b}]: How does the user's privacy-protecting behavior adapt with a privacy screen filter in public settings?
    \item[\textbf{RQ2 }]: What usability challenges and concerns do users report after experiencing a camera slider in public spaces?
\end{itemize}

\begin{figure*}[!ht]
    \centering
    \includegraphics[width=\linewidth]{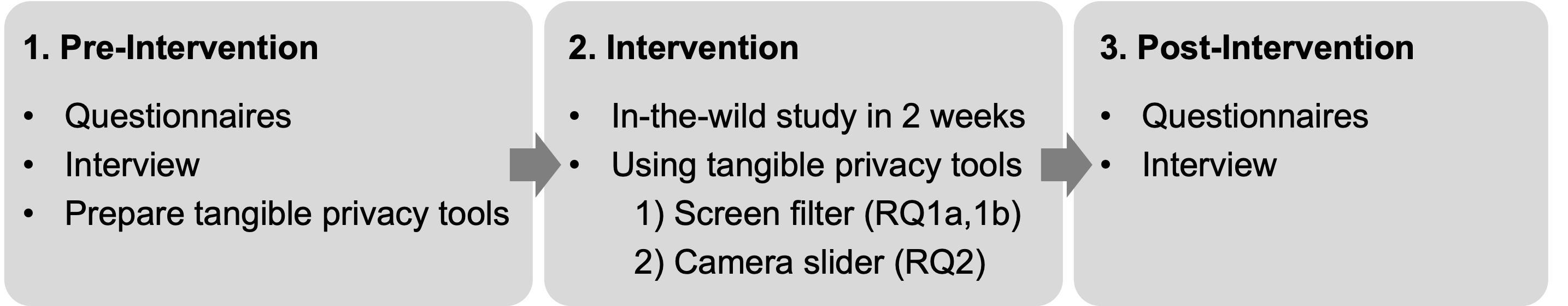}
    \caption{The three phases for our study; pre-intervention, intervention, and post-intervention. }
    \label{fig:study_procedure}
\end{figure*}

\subsection{Participants and Ethics}\label{sec:recruitment_etichs}
We recruited ($N=22$) participants through convenience sampling by approaching individuals at the University of Bergen. We invited individuals to take part in a study about privacy in relation to technology use, with a brief purpose of study section where the specific use of tangible privacy tools was intentionally withheld until after the initial interview to protect the integrity of the pre-intervention data. A total of 24 individuals enrolled in the study, of whom 22 completed it (11 female, 11 male) and two withdrew after the pre-intervention interview. Twenty participants were in the group age of 20-25, one was between 26-30 and one over 30.
With regards to ethical considerations, the study adhered to all relevant ethical guidelines for research involving human participants. We provided and obtained informed consent from all participants, with study procedures and data handling practices clearly communicated including confidentiality and anonymity. The specific procedures and practices included the requirements to participate, such as use a screen filter and camera slider on the smartphone for two weeks, the voluntary participation and right to withdraw and how the processing of personal data would be handled according to the GDPR. The study was registered in the University of Bergen internal system for data protection assessments, which formally documents data collection and handling in line with institutional, ethical, and legal standards for research involving human subjects. The study was conducted in Norwegian and translated to English.

\subsection{Procedure and Study Design}\label{sec:procedure}
The study followed a three-phase protocol: pre-intervention, intervention, and post-intervention (see Figure~\ref{fig:study_procedure}). In the 1) pre-intervention phase, participants signed an informed consent form, which included a purposefully vague description of the study’s aims to minimize response bias while still complying with transparency requirements of the GDPR. Each participant then took part in a semi-structured interview to explore their starting (baseline) attitudes toward privacy and contextual experiences related to smartphone use. Following the interview, they completed a pre-intervention questionnaire and were given the two tangible privacy tools, along with a brief explanation on their application and use.
The participants received the tangible privacy tools (see ~\autoref{fig:tools}) at the end of the pre-intervention phase. We briefly explained the intended use and application of each tool. Following the experimenter's instructions, the participants added the filter and camera lens slider so that the slider was on top of the filter.
During the 2) intervention phase, participants used the tools in the course of their daily lives for a period of two weeks. In order to reduce performance bias and maintain ecological validity, there were no monitoring instructions during this period.
In the 3) post-intervention phase, participants took part in a second semi-structured interview focused on perceived changes in comfort, behavior, social awareness, and the usability of the tools. They also completed a post-intervention questionnaire reflecting on their experiences throughout the intervention period. 

This exploratory study employed a mixed-methods experimental design to explore how tangible privacy tools affect users’ perceptions and behaviors in everyday public physical settings, as well as  usability and social concerns. By combining qualitative interviews with quantitative measures, the study aimed to capture both what changes occurred and why those changes may have taken place during the intervention period. The experimental setup followed a pre–post retention structure, in which participants’ perceptions and behavior were compared before and after a two-week intervention. During this period, the participants used two tangible privacy tools on their smartphones. 
Perceived privacy and behavioral adaptation with the screen filter were evaluated through participants’ reported experiences in the questionnaire and interviews across various public and social settings regarding RQ1a and RQ1b, while the usability and social challenges of the camera slider were evaluated through the interviews for RQ2.
Participants used their own devices throughout, and the within-subjects design allowed each participant to serve as their own control, increasing internal validity by enabling attribution of observed changes to the use of the privacy tools.

\subsection{Instruments: Questionnaire, Interviews}\label{sec:questionnaire}
To assess both subjective experience and behavioral changes, the data collection combined interviews and questionnaire.

\paragraph{Questionnaire}
Participants answered a 10-Likert scale and completed a multiple choice questionnaire both before and after intervention to assess changes in privacy perception and their behavior associated with the use of the privacy screen filter. To ensure alignment with the research questions, questionnaire items were organized into thematic subscales that captured different attitudes of the user experience (e.g., general privacy experiences, perceived privacy contexts, privacy perception, behavioral adaptations). This structure enabled targeted analysis of both perceptual shifts and behavioral adaptations across the intervention period. Both questionnaires were completed independently and anonymously. Phase 1 included the interview to establish initial privacy attitude measures (\autoref{tab:attitudes} in Appendix A.1), and the pre-intervention questionnaire that can be seen together with the post-intervention questionnaire (\autoref{tab:common} in Appendix A.2) was completed after the final interview. Our statistical procedures followed the recommendations in~\cite{tangmisuse,schrum2023concerning}.
To evaluate the entry and exit questionnaire interventions, we performed both a paired t-test and the Wilcoxon signed-rank test. Both are suitable for comparing paired or repeated measures considering a within-subject study. The paired t-test assumes that the data are normally distributed, while the Wilcoxon signed-rank test, a non-parametric alternative, was employed to account for the ordinal nature of the 10-point Likert scale and to provide robustness in the absence of normality.

\paragraph{Interview}
After the questionnaire, two interviews were employed pre and post intervention.
The pre-intervention interview contained six questions and can be fully viewed in Appendix~\ref{appendix_pre_interview}. Participants answered two general questions about their privacy protection while in public, two questions about their previous experience and awareness of shoulder surfing, and two about decisions they had made to protect their privacy in public. Similarly, in the post-intervention interview of seven questions, (for the interview, see Appendix~\ref{appendix_post_interview}), the goal was to extract insights into the participants' experiences and interactions during the two weeks of tangible privacy tool usage. The first two authors conducted the thematic analysis of the interview transcript data by following the procedure outlined in Braun and Clarke~\cite{braun2006using}. Specifically, we 1) entirely read the interview transcripts, 2) highlighted sentences or blocks of text pertaining to our research questions, 3) assigned codes to the identified excerpts, 4) grouped the codes into overarching themes and 5) reviewed the themes and their associated excerpts to ensure consistency and coherence. 

\subsection{Materials and Tools}\label{sec:material_tools}
\begin{figure}[!ht]
    \centering
    \includegraphics[width=\linewidth]{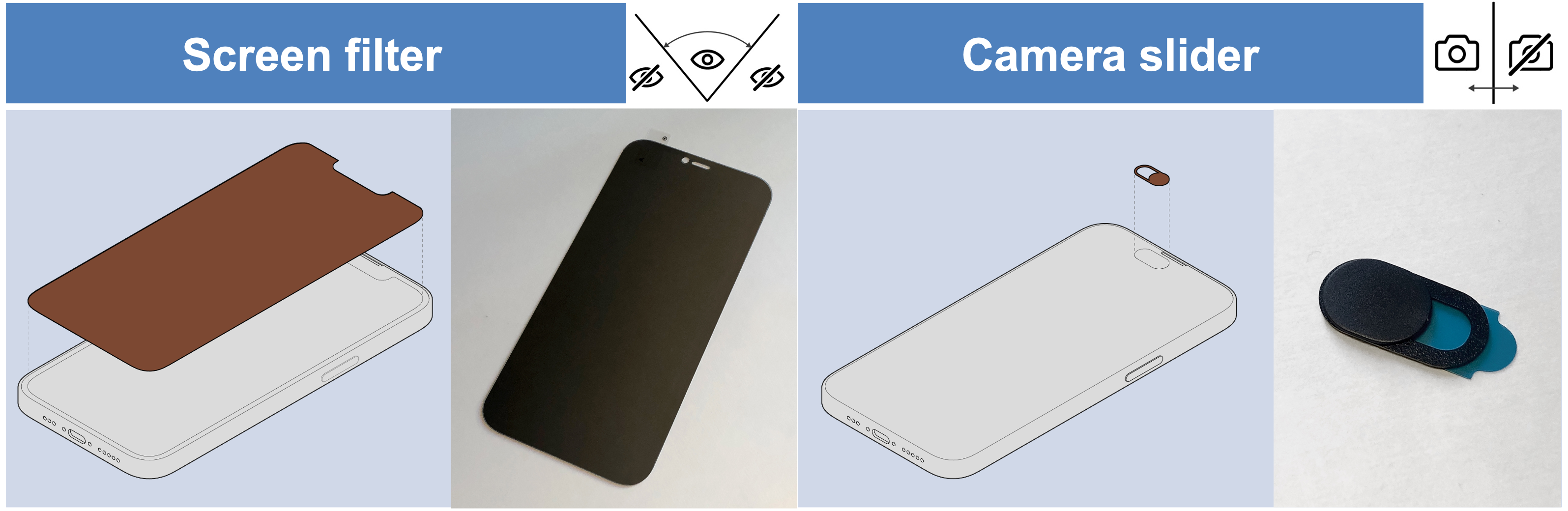}
    \caption{Tangible privacy tools provided to the participants of our study.}
    \label{fig:tools}
\end{figure}

Two tangible privacy tools were used in the study (see~\autoref{fig:tools}): a privacy filter and a camera slider. A privacy filter is a thin, polarized film that can be applied to the smartphone screen to restrict visibility from side angles, enhancing visual discretion in public spaces. The filter used in the study was an off-the-shelf two-way filter. It was designed to limit visibility from lateral (left-right) angles estimated at approximately 60-90 degrees (30-45 degrees from center on each side). It also included openings for the front-facing camera lens to avoid interference with the camera. We prepared three types of filters for Apple iPhone 11, 12/13, and 14/15 devices, and gave them to participants. All participants used their own devices; while 20 participants had Apple iPhone devices, two had Android devices. Since these Android devices' screen dimensions closely matched the iPhone models, their users also could use the filters on them. Several participants already had a protective film on their device (e.g., standard films not explicitly designed for privacy purposes). In these cases, the participants were not required to remove these first. Preliminary testing by multiple participants confirmed that the combination of protective film and privacy filter did not interfere with or significatively reduce touch sensitivity or compromise its function. 
The camera slider (of the brand Plexgear Webcamera Protector) was approximately 18×11×9mm. This was a black plastic mechanism consisting of a frame and circular piece that could be manually slid back and forth to block or expose the lens, enabling users to physically control camera access and limit the potential for visual surveillance. It was to be placed over the front-facing camera of the phone on top of the privacy filter, mounted with self-adhesive tape. 

\section{Results}\label{sec:results}
In this section, we present our analysis procedures and we outline our results from the semi-structured interviews (RQ1a, RQ1b and RQ2) and the entry and exit questionnaire (RQ1a and RQ1b). In Table~\ref{tab:contextual_integrity}, we provide the concepts of the CI adapted to our study context.

\begin{table*}[!ht]
    \centering
    \small
    \caption{Contextual Integrity (CI) Parameters across the two Privacy Tools adapted to the public context.}
    \begin{tabular}{|p{2cm}|p{6.5cm}|p{6.5cm}|}
        \hline
        \textbf{CI Parameter} & \textbf{Shoulder Surfing Context} & \textbf{Unauthorized Camera Access Context} \\
        \hline
        \textbf{Data Subject} & Smartphone user. & Smartphone user. \\
        \hline
        \textbf{Sender} & Individual actively displaying or interacting with on-screen content. & Individual operating the smartphone or app, or recording video/images. \\
        \hline
        \textbf{Receiver} & \textit{Illegitimate}: bystanders or shoulder surfers. \textit{Intended}: legitimate user or chosen peer. & \textit{Illegitimate}: remote party, attacker, or overprivileged system process/application accessing visual data. \textit{Intended}: legitimate user or application with explicitly granted camera access. \\
        \hline
        \textbf{Attribute} & Sensitive on-screen visual content (e.g., banking details, messages, or dating app interactions). & Visual data (images or video) captured by the front camera, potentially revealing personal surroundings or identity. \\
        \hline
        \begin{tabular}[t]{@{}l@{}}\textbf{Transmission}\\\textbf{Principle}\end{tabular} & Expectation that on-screen content in public is visually private and not observed by others. & Expectation that camera activation and data capture occur only under explicit, intentional user consent. \\
        \hline
    \end{tabular}
    \label{tab:contextual_integrity}
\end{table*}

\begin{table*}[!ht]
    \caption{Descriptive statistics - participants' privacy attitudes in the pre-intervention phase. For all the questions the scale is: 1 = Never, 10 = Always, except for the fifth one that used 1 = Not important at all, 10 = Very important.}
    \centering
    \small
    \begin{tabular}{|p{11cm}|c|c|c|c|}
        \hline
        \textbf{Question} & \textbf{Mean} & \textbf{SD} & \textbf{Median} & \textbf{Mode} \\
        \hline
        How often do you actively think about who can see your screen when using your phone in a public space? & 5.36 & 2.15 & 5.5 & 7 \\
        \hline
        How often do you do something to cover your screen in order to feel more comfortable in public spaces? & 4.68 & 1.91 & 4.5 & 3 \\
        \hline
        How often have you chosen to delay or avoid using your phone due to privacy concerns in a public space? & 5.27 & 2.16 & 5.5 & 8 \\
        \hline
        Do you change how you use your smartphone depending on who is nearby? & 6.91 & 2.33 & 7 & 7 \\
        \hline
        How important is it to you that the information on your phone is only shared with people you choose yourself? & 7.60 & 2.15 & 7.5 & 10 \\
        \hline
        When you use your phone in social situations, do you tend to feel self-conscious about what’s on your screen? & 6.54 & 2.09 & 7 & 8 \\
        \hline
        When you use your phone in social situations, do you tend to change how you use your phone depending on who is around you? & 6.50 & 2.48 & 7 & 8 \\
        \hline
        \end{tabular}
    \label{tab:baseline}
\end{table*}

\begin{table*}[!ht]
    \caption{Top three ranked categories across different smartphone use contexts (choices: library, lecture hall, common areas, street/park, café, study hall, transport, gym) and data sensitivity (choices: text/conversations, images/videos, banking/financial info, social media, study/work info, notes/calendar, shopping info.}
    \centering
    \small
    \renewcommand{\arraystretch}{1.3}
    \begin{tabular}{|p{4.25cm}|p{3cm}|p{3cm}|p{3cm}|}
        \hline
        \textbf{Category} & \textbf{1st Rank} & \textbf{2nd Rank} & \textbf{3rd Rank} \\ \hline
        Screen Content Sensitivity & Banking / Financial Info & Texts / Conversations & Images and Videos \\ \hline
        Highest Frequency Context & Common Areas & Public Transport & Gym \\
        \hline
        Most Comfortable Use Context & Library & Common Areas & Street / Park \\
        \hline
    \end{tabular}
    \label{tab:ranking_summary}
\end{table*}

\subsection{Questionnaire Results}\label{sec:quantitative_results}
Given $n=22$ participants and the use of a 10-point Likert scale, which is an ordinal scale, we ran both a paired t-test and the Wilcoxon signed-rank test to ensure the validity of our findings across different statistical assumptions~\cite{schrum2020four}.
We provide the p-value of the single questionnaire item comparisons to explain the statistical significance and the related effect size that provides the magnitude of the effects. 

Overall, to establish internal consistency, the Cronbach’s alpha ($\alpha$) values for the questionnaire subscales indicated acceptable to good reliability. The privacy perception subscale demonstrated acceptable internal consistency ($\alpha \geq 0.70$)~\cite{lance2006sources} in the pre-intervention, while dropping in the post-intervention. In contrast, the behavioral adaptation subscale was slightly below the threshold at pre-intervention but reached an acceptable level post-intervention. In the pre-intervention questionnaire, alpha coefficients were $0.80$ for the \textit{privacy perception} subscale and $0.66$ for the \textit{behavioral adaptation} subscale. For the post-intervention questionnaire, reliability estimates ranged from $0.64$ for the \textit{privacy perception} subscale to $0.76$ for \textit{behavioral adaptation}. Subscales that fell below the criterion suggest potential for refinement, although these values should be interpreted cautiously given the limited sample size.

\subsubsection*{Initial Privacy Attitudes}\label{sec:initial_attitudes}
At the start of our investigation, participants demonstrated awareness and concern regarding privacy when using smartphones in public (see~\autoref{tab:baseline}). On average, they sometimes actively thought about who could see their screen (Mean = 5.36, Median = 5.5) and often chose to delay or avoid phone use due to privacy concerns (Mean = 5.27, Median = 5.5). Participants also adapted their phone use reliably depending on social proximity (Mean = 6.91, Median = 7). Furthermore, the importance placed on selective information sharing was high (Mean = 7.60, Median = 7.5). However, physical protective actions, such as covering the screen, were less frequent (Mean = 4.68, Median = 4.5). Among our participants, only one had previously used a privacy screen filter before, and approximately half reported generally covering their phone cameras (e.g., not necessarily using dedicated camera sliders).
Table~\ref{tab:ranking_summary} further contextualizes these attitudes by identifying smartphone use patterns, where the so-called attribute in the CI framework considered were ranked. Banking and financial information were higher ranked in the screen content sensitivity, underscoring users' heightened concern for the visibility of personal and financial data in public physical settings, followed by texts and conversations, then by images and videos. Common areas and public transport were the most frequent contexts for phone use, while participants felt most comfortable using their phones in solitary or semi-social spaces like libraries and common areas.

\begin{figure}[!ht]
    \centering
    \begin{subfigure}{0.475\linewidth}
        \centering
        \includegraphics[width=\linewidth]{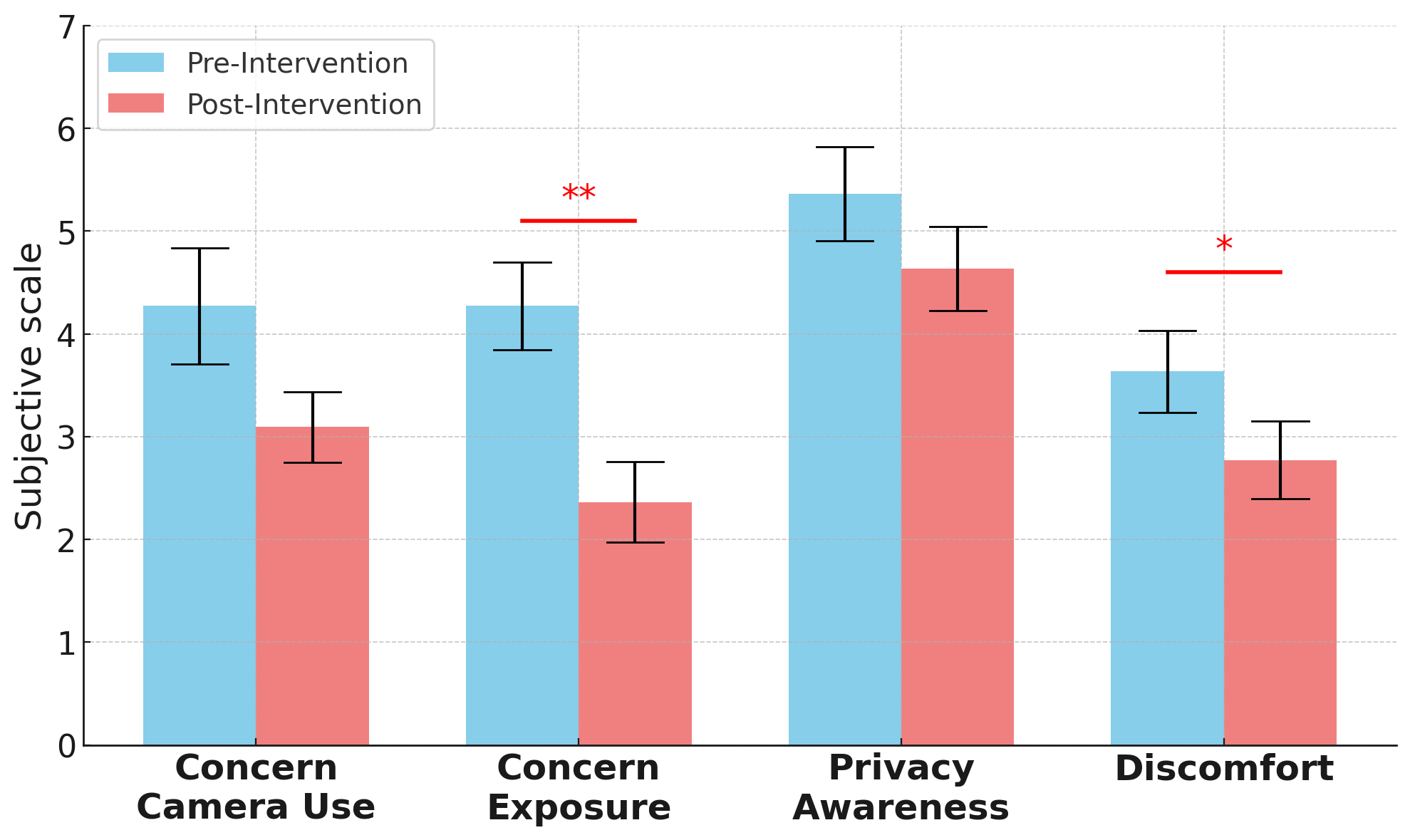}
        \caption{Privacy Perception.}
        \label{fig:privacy-perception}
    \end{subfigure}
    \begin{subfigure}{0.475\linewidth}
        \centering
        \includegraphics[width=\linewidth]{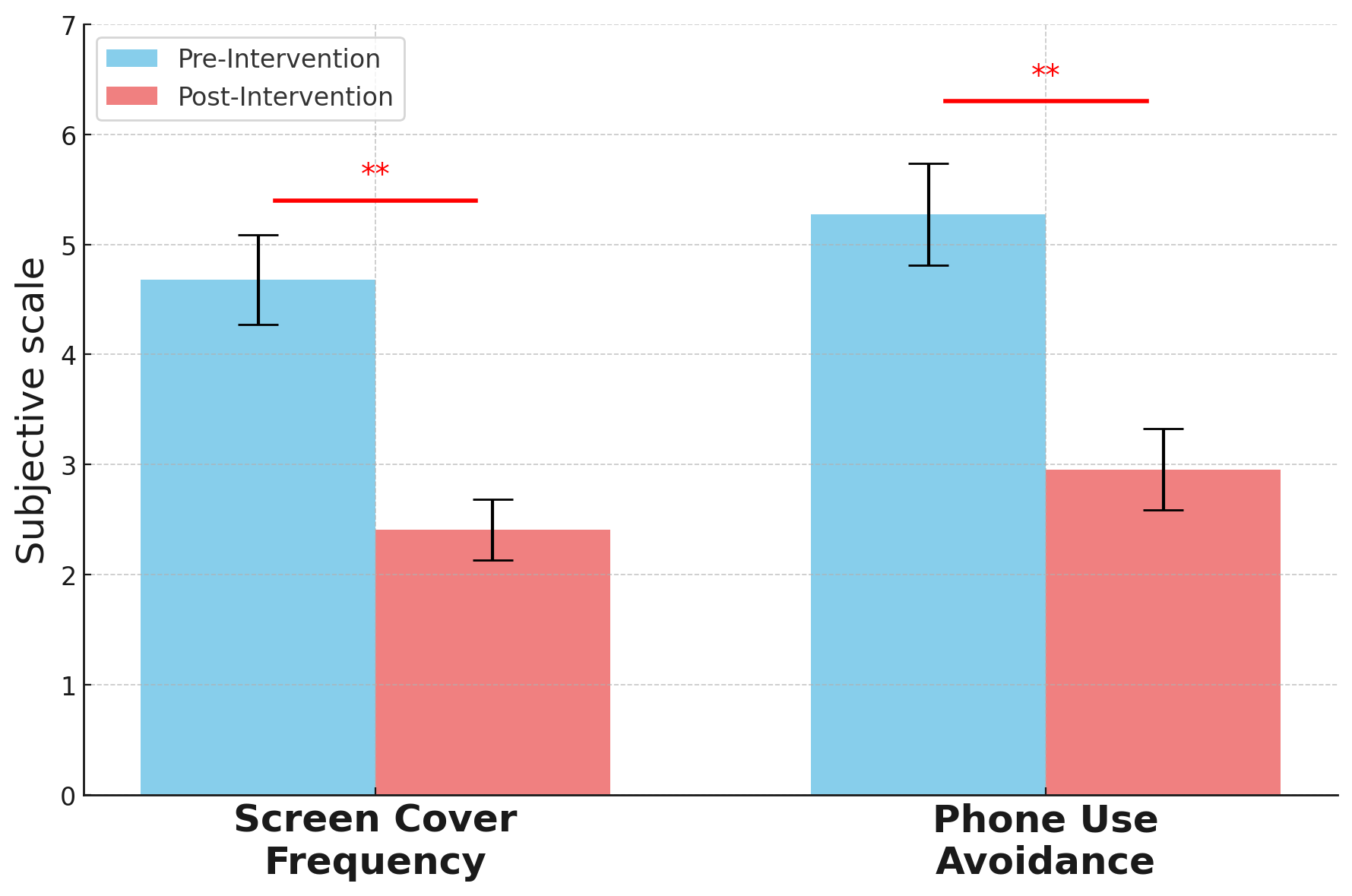}
        \caption{Behavioral Adaptation.}
        \label{fig:behavioral-adaptation}
    \end{subfigure}
    \caption{Privacy Perception - Behavioral Adaptation *: $p < 0.05$, **: $p < 0.01$. Original questionnaires were 10-Likert scale and we cut above 7 point of graph for readability. }
    \label{fig:privacy-vs-behavior}
\end{figure}

\subsubsection*{\textbf{RQ1a} Privacy Perception}\label{sec:privacy_perception_results}
\begin{figure}[!ht]
    \centering
    \includegraphics[width=0.7\linewidth]{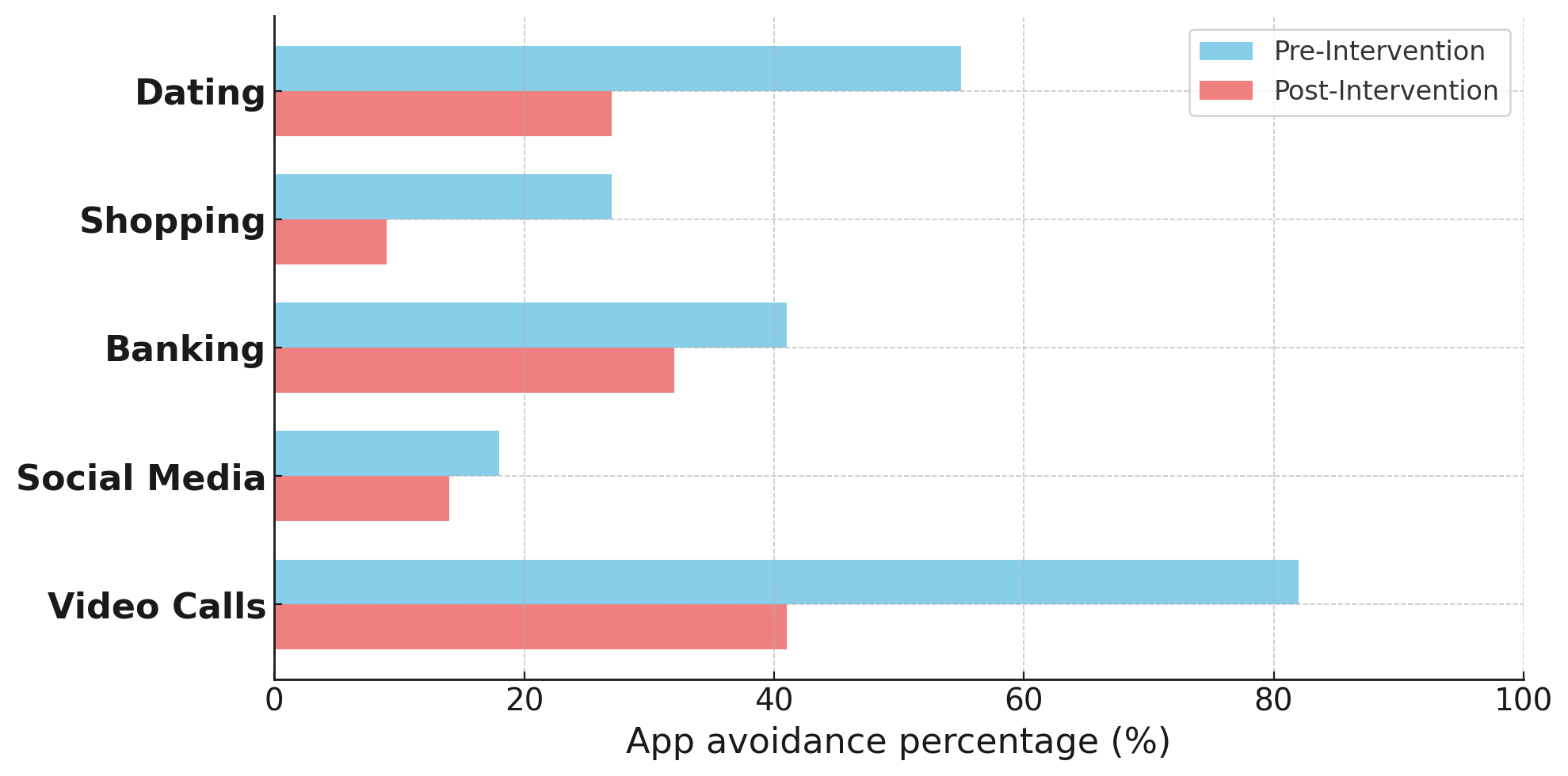}
    \caption{Percentage differences per app avoidance usage reported in pre (blue) and post (red) interventions.}
    \label{fig:app_type_avoidance}
\end{figure}

Following the two-week intervention, participants demonstrated moderate but meaningful improvements in their concerns, awareness and discomfort (see ~\autoref{fig:privacy-perception}). The frequency of discomfort during smartphone use in public decreased significantly ($p = 0.034, d = 0.41$), suggesting increased confidence and comfort. Notably, concern about unwanted screen exposure dropped ($p < 0.001, d = 1.1$), indicating heightened perceived control over privacy risks in public physical settings.
While concern about unauthorized camera use showed a decreasing trend, it did not reach statistical significance, and awareness of screen visibility remained largely unchanged $(p > 0.05)$, reflecting sustained vigilance despite reduced anxiety. Although sensitivity to specific content types, such as texts, banking information, social media, images, and study/work-related content, remained stable with no statistically significant changes (see~\autoref{tab:results}), participants, nonetheless, reported reduced avoidance of several app categories in public physical settings (Figure~\ref{fig:app_type_avoidance}). The proportion of participants who avoided video calls dropped substantially from 82\% pre-intervention to 41\% post-intervention. Avoidance of social media apps declined slightly, from 18\% to 14\%. More pronounced reductions were observed for shopping apps from 27\% to 9\% and dating apps from 55\% to 27\%, suggesting a shift in comfort even in the absence of measurable changes in perceived content sensitivity.
Overall, these findings reveal that the intervention effectively reduced generalized privacy concerns and discomfort while maintaining caution toward privacy risks, indicating a positive recalibration of perceived privacy management in public smartphone use also reflecting in CI attributes including types of information and the public context where the norms governing their transmission.

\subsubsection*{\textbf{RQ1b} Behavioral Adaptation}\label{sec:behavioral_adaptation_results}
The intervention prompted indicated behavioral adaptations toward enhanced privacy protection in public smartphone use (see ~\autoref{fig:behavioral-adaptation} and ~\autoref{tab:results} below). The frequency of physically covering the screen decreased significantly post-intervention ($p < 0.001, d = 1.1$), suggesting that participants felt less need for explicit physical protective behaviors, potentially due to increased confidence in more subtle or effective strategies.
At the same time, reported avoidance of smartphone use due to privacy concerns declined markedly ($p < 0.001, d = 1.13$), indicating a broader behavioral shift toward engaging more openly with their devices in public without compromising privacy. 
Context-specific comfort increased significantly in more privacy-sensitive environments, with participants reporting greater ease using their phones in lecture halls ($p = 0.005, d = 0.74$), public transport ($p = 0.01, d = 0.67$), and gyms ($p = 0.042, d = 0.49$). No significant changes were observed in other common contexts (e.g., common areas, streets, cafés), suggesting the intervention’s perceptual impact was strongest in settings where users initially felt more vulnerable. 
These exploratory findings indicate that participants became more comfortable accessing a wider range of smartphone functionalities in public spaces. The screen filter successfully re-established the transmission principle against shoulder surfing. This is evident in the data showing participants felt significantly less need for explicit behavioral guarding (e.g., physical covering of the screen) and were willing to access sensitive attributes (banking, dating apps) in vulnerable contexts like public transport. This shift confirms the tool successfully enforced the intended contextual norm for information flow. Together, the results for RQ1b demonstrate that the intervention not only influenced participants’ attitudes but also led to meaningful shifts in reported app usage patterns, highlighting a transition from privacy concern to adaptive smartphone use when using the tangible privacy tools. 

\begin{table*}[!ht]
    \centering
    \caption{Privacy perception and behavioural adaptation comparison between pre and post interventions. Results from paired-sample t-tests (t) and Wilcoxon signed-rank tests (w) are shown. Statistically significant results are highlighted in bold ($*p < 0.05$, $** p < 0.01$).}
    \small
    \renewcommand{\arraystretch}{1.1}
    \begin{tabular}{|l|l|c|c|}
    \hline
    \textbf{} & \textbf{Item} & \textbf{p-value} & \textbf{Effect Size} \\
    \hline
    \multirow{7}{*}{\textbf{Privacy Perception - Sensitivity}}
        & Texts/Calls & t = 1.000, w = 0.786 & 0.000 \\
        & Banking Info & t = 0.847, w = 0.887 & -0.042 \\
        & Pictures/Videos & t = 0.919, w = 0.894 & 0.022 \\
        & Social Media & t = 0.584, w = 0.701 & -0.119 \\
        & Study/Work Info & t = 0.136, w = 0.062 & -0.331 \\
        & Notes/Calendar & t = 0.949, w = 0.810 & 0.014 \\
        & Shopping Info & t = 0.657, w = 0.520 & -0.096 \\
    \hline
    \multirow{8}{*}{\textbf{Behavioural Adaptation - Context}}
        & Library & t = 0.776, w = 0.908 & -0.062 \\
        & \textbf{Lecture Hall} & \textbf{t = 0.002**, w = 0.005**} & \textbf{0.741} \\
        & Common Areas & t = 0.247, w = 0.251 & 0.254 \\
        & Street/Park & t = 0.248, w = 0.125 & 0.253 \\
        & Café & t = 0.329, w = 0.246 & 0.213 \\
        & Study Hall & t = 0.192, w = 0.191 & 0.288 \\
        & \textbf{Transport} & \textbf{t = 0.005**, w = 0.010*} & \textbf{0.669} \\
        & \textbf{Gym} & \textbf{t = 0.036*, w = 0.042*} & \textbf{0.492} \\
    \hline
    \end{tabular}
    \label{tab:results}
\end{table*}

\subsection{Thematic Analysis of User Interview}\label{sec:qualitative_results}
As noted in Kumar et al.~\cite{kumar2024roadmap}, qualitative reports using applied CI analysis can enhance the socio-political aspects into the more common quantitative methods. We followed their recommendations while reporting the interview results in this section, in particular identifying CI parameters, contextualizing data - in our case category of applications - and expectations, interpreting meaning, experience and social context. Participants engaged readily with the interview format. Through thematic analysis of post-intervention semi-structured interviews we gained insights into participants' experiences, from which emerged nine codes which could be grouped into three themes: (i) perceived privacy, (ii) behavioral adaptation, and (iii) adoption and usage barriers. Thematic saturation was observed across the themes to mitigate reporting bias with a strong Cohen's kappa level of agreement~\cite{mchugh2012interrater} between the coders of $0.83$.

\begin{table}[!ht]
    \caption{Themes and codes from the thematic analysis.}
    \label{tab:qualitative_results}
    \centering
    \small
    \begin{tabular}{|cl|}
        \hline
        \textbf{Themes} & \textbf{Codes} \\ \hline
        \multirow{3}{*}{\textbf{Perceived Privacy}} & Consciousness and Privacy Comfort \\
         & Selective App Engagement \\
         & Distrust in Privacy Tool Effectiveness \\ \hline
        \multirow{3}{*}{\textbf{Behavioral Adaptation}} & Reduced Behavioral Guarding \\
         & Expanded Public Device Usage \\
         & Social Interaction Friction \\ \hline
        \multirow{3}{*}{\textbf{Adoption and Usage Barriers}} & Technical Usability Barriers \\
         & Social Signalling Concerns \\ 
         & Emergent Privacy Awareness \\ \hline
    \end{tabular}
\end{table}

\subsubsection{\textbf{RQ1a - Perceived Privacy}} 
\paragraph*{Consciousness and Privacy Comfort}
Seventeen participants out of twenty-two reported feeling more aware of their screen’s visibility to others in public physical settings. The use of privacy tools appeared to increase their consciousness around what others could potentially see, making them feel more in control of their digital privacy. For example, participant P05 said \textit{“It just felt a bit more private, that was nice.”} Similarly, participant P11 reflected, \textit{“I’ve thought a lot about it when I’ve been in public, that people don’t see the same from other angles. That has felt a bit comforting to me."}. With the added sense of protection, they described feeling less worried about who might be looking at their screens. Participant P04 said, \textit{“You think less about it... just use it”}. Participant P15 shared, \textit{“But then I noticed I was like...‘Wait, I have that screen on,’ so I did things because of that, instead of not using it”}. Three participants (P06, P10 and P15) used the filters implicitly, indeed P06 \textit{"I was conscious of the camera lens, but wasn't conscious of the filter, [...] it's a good thing that they're just doing their job without me thinking about them."}. P14 added, \textit{"I started to be a bit more conscious. I think about it once in a while. 'Oh yes, I have this screen filter.' I think I've just kind of forgotten when I've got it, so it hasn't really changed the way I've used my phone."}.

\paragraph*{Selective App Engagement}
The tools influenced participants’ willingness to engage with certain apps in public. Fourteen of them described choosing to open or avoid specific applications based on how protected they felt—especially apps that could reveal personal or sensitive content. In particular, participant P03 directly included attribute and transmission principle, two aspects of the CI framework, in selecting app and context, they shared \textit{“Then I could use  [dating app] at the gym”.} Likewise, participant P24 highlighted the attribute (e.g., social media app) by explaining, \textit{“When I’m on [social media], you kind of don’t want people to see it. [...] I didn’t feel so exposed”.}. Two participants (P15 and P22) don't feel constrained in using sensitive information related to banking, personal interactions or social media while in public transport, P15: \textit{"In the bus I was like, 'Wait, I have on that screen filter', so I did things instead of not using it. So then the bank, pictures, messages and things like that"}. This shift means that users realigned their utilization of sensitive data with the attribute parameter of CI.

\paragraph*{Distrust in Privacy Tool Effectiveness}
Participants (n=3) expressed doubts about the actual effectiveness of the tools in blocking screen visibility. This skepticism often originated from the fact that they could still see the screen clearly themselves, which made it difficult to fully trust that others could not do the same from certain angles. Participant P03 and P17 discussed about social media within the family context, with P03 stating that \textit{“I was around my family and I've gotten a [video/image from a social media app] from a friend and I didn’t really trust the screen either”}. Participant P21 elaborated, \textit{“I haven’t felt that others can’t see my screen. It’s just that when I look at it from the front, I can see it myself. So, like, I don’t think my brain really feels it’s private".}

\subsubsection{\textbf{RQ1b - Behavioral Adaptation}}
\paragraph*{Reduced Behavioral Guarding}
Nine participants reported that they stopped engaging in typical privacy-protective behaviors, such as turning down screen brightness or angling their body to shield the display. The perceived protection provided by the tools seemed to replace these earlier habits, reflecting a transfer of responsibility from the user to the tool. Participant P20 remarked, \textit{“I don’t turn down the brightness and stuff like I normally do automatically”}. Participant P16 added, \textit{“I think I’m a bit more open in my body language. That’s the only thing I’ve noticed”}. Participant P15 noted, \textit{"you quickly get used to the idea that if you want to show things, you have to hand the phone over, so they can hold it themselves or try to angle it straight towards people. I feel like that became automatic very quickly [...] it was more of an adaptation thing."}

\paragraph*{Expanded Public Device Usage}
Ten participants noted a greater willingness to use their smartphones in public places, particularly in public transport. Participant P07 include useful information regarding the CI receiver: \textit{"I write a message to my mother without feeling that someone else will get it"}. Participants P08, P10, P16, P17, P18, P20 in the public transport felt safer and more aware, with P08 stating \textit{"I was checking something sensitive while sitting on the tram, then I felt a little bit safer."}.

\paragraph*{Social Interaction Friction}
While the tools enhanced individual privacy, they also introduced challenges in social contexts. Participants (n=5) reported that it became harder to share content with others, such as showing photos or videos to friends. Participant P24 noted, “\textit{Another time I was reminded by others commenting on the screen [...] like...‘Okay, I can’t see... Why is your screen black?’}”. Participant P17 added, \textit{“Sometimes when you want to show things to people, you have to find the right angle”}. Participant P07 pointed the problem of showing videos to friends \textit{"I try to show the same video to three different people, I have to show it several times so that everyone can see it."}.

\subsubsection{\textbf{RQ2 - Adoption and Usage Barriers}}
\paragraph*{Usability Barriers}
Usability issues were widespread and frequently mentioned. Participants (n=11) reported frustration with reduced touch sensitivity, facial recognition unlocking malfunctions, and a general decrease in product responsiveness. Participants P08, P12, P13, P14, P20, P21 reported problems with the FaceID and P12 stated, \textit{“Yeah, it was a bit more annoying, with Face ID not working”}, with P13 that extended the problem related to the camera, \textit{"And then it's not so easy to FaceTime, I use it several times a week"}. Furthermore, Participants P15 and P17 found problems in taking pictures, \textit{"it's affected the quality of my front camera"}. Unfortunately, we also detected from the interviews that several participants had problems in maintaining the camera slider attached to their smartphone. Such usability barriers disrupted the normal, expected function of the smartphone.

\paragraph*{Social Signalling Concerns}
Six participants voiced concerns about how their use of privacy tools might be interpreted by others. The camera slider, in particular, was seen as a possible signal of paranoia, secrecy, or social withdrawal. Participant P08 reflected, \textit{"If others saw the camera protector, they might think I was a kind of anxious person"}. Participant P22 explained, \textit{"Now I have a bit more privacy on my screen, but at the same time, I want to be perceived as an open person, and then it feels wrong"}. Participant P03 discussed with others and stated \textit{"Over the camera, I've talked to several people who think that I'm a bit paranoid, but we also agreed that it's smart"}. Participant P14 discussed about the comments received,  such as \textit{"Curiosity or maybe a bit judgmental. Like 'Do you have it? Why? Are you so super paranoid?"}. The camera slider can address the digital intrusion norm, but introduced a social norm disruption. This illustrates that while the physical tool can potentially secured the CI principle against illegitimate senders, it conflicted with implicit social norms surrounding openness and trust in public.

\paragraph*{Emergent Privacy Awareness}
Five participants explained that their awareness to protect their privacy became more present in their everyday life by being exposed to our study. Participant P08 reported, \textit{"I maybe got a bit self-conscious about that camera protector"}. Participant P17 stated, \textit{"The fact that you yourself somehow don't think it's possible that someone is looking through your camera, [...] so I become a bit more self-conscious"} and later came back to the topic and indicating a partial re-establishment of the security norm against this abstract threat mentioning that \textit{"I haven't been too worried about people looking through my camera. But it was kind of nice to know that even if someone did it, they couldn't see anything."}. Participant P22 highlighted a difference in terms of effects of the two tools used during the two weeks, in particular \textit{"In terms of privacy, I'm much more afraid of the data collected inside the phone, if you can call it that, than the people who look from the side. [...] Because it's nice to be absolutely sure that it's not possible to look at you through the camera."}
\section{Discussion}\label{sec:discussion}

This paper, focusing on tangible smartphone privacy tools, highlights that the privacy screen filter demonstrated a positive impact on perceived privacy and facilitated behavioral adaptation, whereas the camera slider encountered more social criticism and usability difficulties. In this two-week in-the-wild study with N=22 participants employed questionnaires and interviews, we found: 

\indent \textbf{1. The privacy screen filter enhanced perceived privacy and encouraged behavioral adaptation in public settings (RQ1a–RQ1b):} participants reported lower concern about unwanted screen exposure, decreased physical guarding behaviors, and greater comfort using sensitive apps (e.g., banking, dating) in shared environments. These findings show how the screen filter re-established appropriate information norms under the CI framework.

\indent \textbf{2. The camera slider’s effectiveness was constrained by usability issues and social perception (RQ2):} interference with functions such as FaceID and camera use, together with worries about appearing anxious or paranoid, hindered adoption and social acceptability, limiting its perceived privacy benefits.

\indent \textbf{3. Perceived proximity of privacy threats may shape interaction with tangible privacy tools:} while still exploratory, in our study, participants reacted more strongly to concrete, immediate risks (e.g., shoulder surfing) than to abstract ones (e.g., remote camera access). This emphasizes the designing of tools that address both technical, psychological and social dimensions of privacy.

\subsection{Privacy Screen Filter}
In this study, we applied the CI framework. A privacy screen filter therefore helps preserve CI by enforcing the appropriate transmission principle, limiting the flow of information so that it reaches only the intended recipient and preventing passive data acquisition by illegitimate recipients (e.g., shoulder surfers). 

\paragraph{Enhancing Privacy Perception}
The privacy screen filter significantly improved perceived screen privacy (RQ1a), evidenced by a notable decrease in concern about unwanted screen exposure $(p < 0.001, d = 1.1)$ and reduced discomfort during public smartphone use $(p = 0.034, d = 0.41)$. Participants frequently reported feeling more aware of their screen's visibility to others in public physical settings, which made them feel more in control of their privacy. Furthermore, the screen filter's effectiveness was sometimes so seamless that participants like P06 noted it was \textit{"a good thing that they’re just doing their job without me thinking about them"}. No statistical significance was found regarding the sensitivity to explicit content information on the smartphone screen. However, participants reported that the filter enabled them to comfortably access a wider range of previously avoided app categories, such as video calls (avoidance decreased substantially from 82\% pre-intervention to 41\% post-intervention), shopping apps (27\% to 9\%), and dating apps (55\% to 27\%). This perception of app usage expansion signifies a realignment of user utilization with the attribute (type of information) parameter of CI. Participants felt comfortable using applications containing sensitive information, like banking or dating apps, in public transport, confirming the findings in~\cite{farzand2022shoulder}, because the screen filter helped establish an appropriate information flow while preventing unauthorized viewing because of shoulder surfing. As Participant P08 stated, \textit{“I was checking something sensitive while sitting on the tram, then I felt a bit safer”}, which supports how the tool facilitated adherence to their personal contextual norms for information sharing.

\paragraph{Adapting Behavior in Public Use}
Our study also noted behavioral adaptations (RQ1b) toward enhanced privacy protection in public smartphone use. The frequency of physically covering the screen decreased significantly post-intervention $(p < 0.001, d = 1.1)$, suggesting that participants felt less need for explicit privacy-protective behaviors, possibly due to increased confidence in the filter's subtle effectiveness. Participant P20 remarked, \textit{“It’s kind of like I don’t turn down the brightness and stuff like I normally do automatically”}. Meanwhile, reported avoidance of smartphone use due to privacy concerns declined $(p < 0.001, d = 1.13)$, indicating a behavioral shift toward expanded public device usage. This behavioral adaptation directly relates to the transmission principle of CI, as users felt more assured that their information was being shared only under their desired conditions and not intercepted by shoulder surfers. The increased comfort was particularly significant in environments previously investigated with similar results in~\cite{farzand2022shoulder}, such as lecture halls $(p = 0.005, d = 0.74)$, public transport $(p = 0.01, d = 0.67)$, and gyms $(p = 0.042, d = 0.49)$, highlighting how the filter helped align information flow with specific contextual norms where screen visibility is often expected to be limited. 
\paragraph{Usability and Tool Effectiveness}
While predominantly positive, the screen filter presented some usability challenges. Participants occasionally (n=2) needed to adjust their viewing positions, especially when not looking at the phone head-on, with participant P08 noting, \textit{"If I’m sitting on the sofa in a few different types of positions and I’m looking at my phone, I notice that I either have to straighten up or hold it"}. Furthermore, social interaction friction was observed when participants (n=5) attempted to share screen content with others. This suggests that while the filter improved individual privacy by controlling the transmission principle for personal viewing, it could inadvertently impede the ease of shared viewing, a different facet of information flow within social contexts. Qualitatively, three participants also expressed a distrust in privacy tool effectiveness, as they could still see their own screen clearly from the front, making it difficult to fully believe others could not do the same from different angle. Given that participants' comfort increased significantly in specific contexts, future screen filters could be designed to integrate with smartphone sensors or app usage data. This could allow for context-aware activation, automatically engaging the filter when the user enters a privacy-sensitive environment or opens a sensitive application (e.g., banking app), aligning with the transmission principle and attribute parameters of CI. Moreover, filters that can dynamically adjust their privacy level based on the proximity of people in front of the screen via a temporary stopping of the "privacy" filter mode, could be a further development opportunity to explore.

\subsection{Camera Slider}
While the screen filter protects against illegitimate visual receivers (e.g., shoulder surfers) in the physical space, the camera slider protects against illegitimate visual senders (e.g., app, attacker, system process) that was not contextually allowed to receive it in the digital space. The camera slider, while intended to prevent unauthorized camera access, faced more social criticism and usability challenges (RQ2). As communicated in the interview results (n=11), its adoption was limited due to issues with the slide, for example four participants reported that its use sometimes interfered with standard phone functions like facial recognition unlocking, or using camera-centric applications, such as some social media, indeed Participant P17 exposed, \textit{it's affected the quality of my front camera}. A solution about FaceID, obfuscating with a physical blur layer the camera, was proposed for FaceID smartphone authentication method achieving 95\% of accuracy for authentication and 100\% for deauthentication~\cite{cardaioli2023blufader}. Furthermore, participants (n=4) expressed concerns about how the visible presence of a camera slider might be perceived by others, leading to worries about appearing anxious or paranoid, Participant P08 explained, \textit{If others saw the camera protector, they might think I was a kind of anxious person}. This social signaling aspect can create inconvenience, potentially stopping the perceived privacy benefits if users feel socially disadvantaged by using the tool. The social learning theory applied to the webcam covering demonstrated that people with low self-control prefer to use the webcam covering when they are aware of the potential risks, or after having experienced privacy violation themselves~\cite{nicholson2025much}. This indicates that while tangible tools can enhance privacy perception, their design should also account for social dynamics, as found for laptop webcam~\cite{machuletz2018webcam}, and practical user experience to achieve broader adoption and behavioral change. Future designs should aim to mitigate these practical inconveniences and social friction points to maximize the effectiveness of such privacy solutions. Integrating automation with clear user feedback may support privacy awareness even when users forget to act, helping maintain trust and reducing social friction. 

\subsection{Contextual Integrity}
Regarding the CI framework, the passive privacy screen filter successfully re-established CI by enforcing the appropriate transmission principle against shoulder surfing. Before the intervention, the norm required the data subject to actively intervene (e.g., angling the screen, covering it) to enforce the non-interception rule (see Table~\ref{tab:baseline}). Limiting the flow of sensitive attributes (e.g., banking or dating app data), as highlighted also during the interview from P15 and P22: "didn't feel so exposed", from illegitimate receivers significantly improve the reduction of physical covering and change the transmission principle. Indeed, Participant P07: "I write a message to my mother without feeling that someone else will get it". The filter reinforces the integrity of the communication channel by guaranteeing the information reaches only the chosen receiver. This reintroduce the expected privacy norm that led to significant behavioral adaptation, since participants dropping physical guarding behaviors and expanding their use of sensitive applications, as reported in the qualitative findings with around the half of participants (n=11), including in public transport and lecture halls, as in~\cite{farzand2022shoulder}, and gym where P03: "could use [dating app] at the gym". In contrast, the camera slider illustrates a conflict in norm preservation: while it enforces a technical CI principle against illegitimate remote senders in the digital space, its physical visibility disrupted critical social norms in the public context. It introduced social friction when users needed to share content with authorized receivers (e.g., friends), forcing users to adjust the angle or hand over the phone. Participants’ worries about being negatively judged by others (n=6) and the presence of technical usability conflicts (n=11) confirm that attempting to enforce one CI norm can lead to the disruption of other practical usability property or social norms.

\subsection{Psychological Distance and Design Implications}
The different findings between the two tools offer implications for the design and deployment of tangible privacy tools. First, our study confirms that usability is relevant for the adoption and effectiveness of privacy solutions~\cite{livingstone2008taking}. The camera slider's technical glitches and social friction points can create problems for the participants, while the screen filter facilitated demonstrable changes in behavior and perception. This suggests that merely providing a privacy mechanism is insufficient; its design should proactively mitigate practical inconveniences and account for social dynamics to enable behavioral changes.

\paragraph{Psychological Distance and Tangible Privacy Perception}
While differences in usability offer a initial explanation for the divergent reception of the two tools, there may also be a complementary psychological dimension at play. Specifically, users may perceive and respond more readily to privacy threats that feel immediate and tangible, such as the risk of shoulder surfing mitigated by the screen filter, compared to more abstract or remote threats, like unauthorized camera access. This distinction aligns with the Construal Level Theory (CLT), which states that individuals mentally establish events that are psychologically distant, whether in time, space, social context, or likelihood, at a higher, more abstract level, than what is taking place for the individual in the current here-and-now~\cite{trope2010construal}. 
Applied to this study, previous experiences may play a role, together with the immediate visual threat of shoulder surfing being psychologically "closer" and more concrete, leading to a more positive recognition of the screen filter. In contrast, the abstract, potential threat of remote camera access might be psychologically "further", leading to less engagement with the camera slider, especially when compounded by usability issues~\cite{williams2016perfect,adams1999users}. Regarding laptop webcam, individuals’ level of concern does not significantly impact their decision to cover it~\cite{machuletz2018webcam}. Moreover, this interpretation aligns with the control paradox regarding the camera as part of smart assistants studied on tangible privacy feedbacks, where participants felt less risks since the decreased ambiguity about the device status~\cite{ahmad2022tangible}.

\paragraph{Design Implications for Tangible Privacy Tools}
Despite these issues, some participants reported an emergent privacy awareness regarding their camera, acknowledging a new sense of security in knowing that remote access would be blocked, as Participant P17: \textit{"I haven't been too worried about people looking through my camera. But it was kind of nice to know that even if someone did it, they couldn't see anything."}. More studies are needed considering that the literature demonstrated how individuals prefer a manual action to activate the protection on their smart home cameras as they feel to be in control even if the cognitive load is high~\cite{shalawadi2024manual}. On the other hand, the automatic protection of laptop cameras showed more reliability and helped to avoid human error~\cite{do2021smart}. The tool itself demands an additional active step for the user, which requires memory while accessing the smartphone functionalities. In our results, we found contextual predispositions when our participants were bothered by the camera slider, in particular while accessing the FaceID for authenticating themselves and while accessing social media where taking photos or videos is an essential functionality. Our preliminary results would motivate the tangible and embodied community to investigate further effective and simple tangible privacy solutions considering contextual properties, privacy perceptions and social acceptance. Future works could include the designing and testing hybrid, context-aware tangible privacy solutions, such as a privacy screen filter that polarized the screen while accessing to banking app or a camera slider that uncover the camera when individuals take the phone from the pocket since it often follows the authentication procedure with facial recognition. Indeed the hybrid camera covering has been confirmed to be effective~\cite{do2021smart}. Furthermore, future works could also explore how social contexts and the presence of bystanders influence the continued use, perceived safety, and social acceptability of tangible privacy tools. In this context, the bystander can be seen as unauthorized receiver or a potential sender under the CI lens. For instance, observational or play-acted scenarios could be designed where a shoulder surfer challenge the participant’s visible use of a camera cover or screen filter. Such settings would help to understand how users negotiate privacy norms when potential observers or threat actors are present. Previous works studies the camera use in public settings including live streaming, from the streamer side who usually have the direct control of the recordings~\cite{wu2023streamers} and the bystanders one who often are unaware of others people recordings~\cite{faklaris2020snapshot}. Bystanders are interested in being part of the streaming process, including visual consent for the recording~\cite{tebbe2025like} and the possibility to block transmission~\cite{bohn2005social}. Measuring the comfort of the bystanders while the streamer's device is equipped with a camera slider or tangible or digital obfuscation techniques could provide further insights regarding their privacy in public settings.
\section{Limitations}\label{sec:limitations}
Conducting an in-the-wild study involves trade-offs, as participants interact with the tools in their natural environments without direct researcher supervision. Because of some attachment problems of the camera slider in the participants' phones reported during the post-intervention interviews, we decided to not include the questions regarding the camera slider behavioral adaptation in the main body of this report, while keeping the perceived privacy information as in Figure~\ref{fig:privacy-perception}. The unreported questions are available in Appendix~\ref{appendix_post_questionnaire} with descriptive statistics.
The sample size of our investigation was relatively small ($N=22$), especially regarding the questionnaire results in Figure~\ref{fig:privacy-vs-behavior} and Table~\ref{tab:results}, which necessitates caution in interpreting and generalizing the findings. Further, our participants were mostly students belonging to a certain age group, which may have also limited the possible range of responses.

\section{Conclusion}\label{sec:conclusion}
This study explored the real-world impact of tangible smartphone privacy tools, specifically privacy screen filters and camera sliders, on users' perceptions and behavioral adaptations in public spaces. Employing a two-week, mixed-method, in-the-wild design with 22 participants, the research provided insights into the efficacy and challenges associated with these privacy solutions. Our findings highlight distinct outcomes for each tool; the privacy screen filter demonstrated a positive impact on users' perceived privacy and facilitated the behavioral adaptation, while the camera slider elicited more social criticism and usability difficulties.

\subsubsection*{Acknowledgements}
We thank the participants for their active engagement during the two weeks of data collection. We thank the anonymized reviewers for their useful comments. This work was partially supported by the Wallenberg AI, Autonomous Systems and Software Program (WASP) funded by the Knut and Alice Wallenberg Foundation. 

\bibliographystyle{unsrt}  
\bibliography{sample-base}

\begin{landscape}
\section{Questionnaires}
\subsection{Initial Privacy Attitudes and Previous Experiences}

    \begin{table}[!ht]
    \vspace{-0.4cm}
    \resizebox{\linewidth}{!}{
    \begin{tabular}{|l|l|l|c|l|}
        \hline
        \textbf{Index} & \textbf{Question} & \textbf{Categories} & \textbf{Type} & \textbf{Scale}\\
        \hline
           \multirow{11}{*}{\textbf{General privacy experiences}}
            & How often do you actively think about who can see your screen when using your phone in a public space? & & Scale & 1 = Never, 10 = Always \\
            \arrayrulecolor{gray!30}\cline{2-5}\arrayrulecolor{black} 
            & How often do you do something to cover your screen to feel more comfortable in public spaces? & & Scale & 1 = Never, 10 = Always \\
            \arrayrulecolor{gray!30}\cline{2-5}\arrayrulecolor{black} 
            & How often have you chosen to postpone or avoid using your phone due to privacy concerns in a public space? & & Scale & 1 = Never, 10 = Always \\
            \arrayrulecolor{gray!30}\cline{2-5}\arrayrulecolor{black} 
            & Do you change the use of your smartphone depending on who is around? & & Scale & 1 = Never, 10 = Always \\
            \arrayrulecolor{gray!30}\cline{2-5}\arrayrulecolor{black} 
            & How important is it to you that the information on your phone is only shared with people you choose? & & Scale & 1 = Not important at all, 10 = Very important \\
            \arrayrulecolor{gray!30}\cline{2-5}\arrayrulecolor{black} 
            & When you use your phone in social situations, do you tend to feel self-conscious about what is on your screen? & & Scale & 1 = Never, 10 = Always \\
            \arrayrulecolor{gray!30}\cline{2-5}\arrayrulecolor{black} 
            & When you use your phone in social situations, do you tend to change how you use your phone depending on who is around you? & & Scale & 1 = Never, 10 = Always \\
            \arrayrulecolor{gray!30}\cline{2-5}\arrayrulecolor{black}
            & Have you previously used a screen filter for privacy in the past? & & Single & 1 = Never in the past, 2 = Once in the past, \\
            & & & choice & 3 = Occasionally, 4 = Regularly, 5 = Always\\
            & Have you ever covered your phone's camera? & & Single & 1= Never, 2 = Rarely, 3 = Sometimes, \\
            & & & choice & 4 = Often, 5 = Always\\
        \hline
           \multirow{21}{*}{\textbf{Privacy perceived contexts}}
            & & Texts/Calls & & \\
            & & Banking info & & \\
            & & Pictures/Videos & & \\
            & In a public space, how sensitive do you consider the following information on your screen? & Social media & Scale & 1 = Not sensitive at all, 10 = Very sensitive \\ 
            & & Study/Work info & & for each category\\
            & & Notes/Calender & & \\
            & & Shopping info & & \\
            \arrayrulecolor{gray!30}\cline{2-5}\arrayrulecolor{black} 
            & & Library & & \\
            & & Lecture hall & & \\
            & & Common area & & \\
            & How often do you use your smartphone in the following social/public settings? & Street or park & Scale & 1 = Never, 10 = Always \\
            & & Coffee shop & & for each category\\
            & & Study room & & \\
            & & Public transport & & \\
            & & Gym & & \\
            \arrayrulecolor{gray!30}\cline{2-5}\arrayrulecolor{black}    
            & & Library & & \\
            & & Lecture hall & & \\
            & & Common area & & \\
            & How comfortable are you using your smartphone in the following settings? & Street or park & Scale & 1 = Less comfortable, 10 = Highly comfortable \\
            & & Coffee shop & & for each category\\
            & & Study room & & \\
            & & Public transport & & \\
            & & Gym & & \\
        \hline
        \end{tabular}
        }
    \caption{General privacy experiences (reported in~\autoref{tab:baseline} and in~\autoref{sec:initial_attitudes}) and privacy perceived contexts (reported in~\autoref{tab:ranking_summary}) at pre-intervention phase}
    \label{tab:attitudes}
\end{table}

\clearpage
\subsection{Questionnaires asked during pre and post interventions.}
    \begin{table}[!ht]
    \vspace{-0.4cm}
    \resizebox{\linewidth}{!}{
    \begin{tabular}{|l|l|l|c|l|}
        \hline
        \textbf{Index} & \textbf{Question} & \textbf{Categories} & \textbf{Type} & \textbf{Scale}\\
        \hline
           \multirow{14}{*}{\textbf{Privacy Perception}}
            & How concerned are you about unauthorized use of your phone’s camera? & & Scale & 1= Not concerned, 10 = Very concerned \\
            \arrayrulecolor{gray!30}\cline{2-5}\arrayrulecolor{black}    
            & How worried are you that people around you can see your screen? & & Scale & 1 = Not worried at all, 10 = Very worried \\
            \arrayrulecolor{gray!30}\cline{2-5}\arrayrulecolor{black}  
            & How often do you actively think about who can see your screen when using your phone in a public space? & & Scale & 1 = Never, 10 = Always \\
            \arrayrulecolor{gray!30}\cline{2-5}\arrayrulecolor{black} 
            & How often have you felt uncomfortable or regretted using your smartphone in a public space due to privacy concerns? & & Scale &  1 = Never, 10 = Always \\
            & Are there any types of activities or apps you avoid using your phone in public spaces? & & Multiple & Video calls, Social media, Banking, \\
            & & & choice & Online shopping, Dating apps\\
            \arrayrulecolor{gray!30}\cline{2-5}\arrayrulecolor{black} 
            & & Texts/Calls & & \\
            & & Banking info & & \\
            & & Pictures/Videos & & \\
            & In a public space, how sensitive do you consider the following information on your screen? & Social media & Scale & 1 = Not sensitive at all, 10 = Very sensitive \\ 
            & & Study/Work info & & for each category\\
            & & Notes/Calender & & \\
            & & Shopping info & & \\
        \hline
            \multirow{10}{*}{\textbf{Behavioural Adaptation}}
            & How often do you take steps to cover your screen to feel more comfortable in public spaces? & & Scale & 1 = Never, 10 = Always \\
            \arrayrulecolor{gray!30}\cline{2-5}\arrayrulecolor{black}  
            & How often have you chosen to postpone or avoid using your phone due to privacy concerns in a public space? & & Scale & 1 = Never, 10 = Always \\
            \arrayrulecolor{gray!30}\cline{2-5}\arrayrulecolor{black} 
            & & Library & & \\
            & & Lecture hall & & \\
            & & Common area & & \\
            & How comfortable are you using your smartphone in the following settings? & Street or park & Scale & 1 = Less comfortable, 10 = Highly comfortable \\
            & & Coffee shop & & for each category\\
            & & Study room & & \\
            & & Public transport & & \\
            & & Gym & & \\
            \arrayrulecolor{gray!30}\cline{2-5}\arrayrulecolor{black}

        \hline
        \end{tabular}
        }
    \caption{Common questionnaires at Pre/Post-Intervention (reported in~\autoref{sec:privacy_perception_results} and in~\autoref{sec:behavioral_adaptation_results})}
    \label{tab:common}
\end{table}
\end{landscape}

\subsection{Continuation of Post-Intervention Questionnaires}\label{appendix_post_questionnaire} 
\begin{enumerate}
    \item How much did you enjoy using the screen filter in general during this period? (Options: 1 = Disliked, 10 = Liked very much) [mean=$6$, sd=$2.7$, median=$6$]
    \item How often did you open/close the camera cover daily during this period? (Options: 1 = Never, 2 = 1-4 times per day, 3 = 5-10 times per day, 4= 11-20 times per day, 5 = More than 20 times per day) [mean=$2.8$, sd=$1.18$, median=$3$]
    \item How much did you enjoy using the camera cover in general during this period? (Options: 1 = Didn't like it at all, 10 = Liked it very much) [mean=$4.5$, sd=$3$, median=$5$]
    \item How did you feel the privacy filter affected the flow of information from your phone to others? (Options: 1 = No impact at all, 10 = Great impact) [removed since "flow of information" was misunderstood by the participants]
    \item Did you find that the privacy tools changed how others reacted to your phone use in public spaces? (Options: 1 = No change at all, 10 = Big change) [removed from the  questionnaire and asked during the interview - see question 4 Appendix~\ref{appendix_post_interview}]
    \item Did the privacy filter make you feel safer when using your phone in public settings? (Options: 1 = Not safer at all, 10 = Very much safer) [mean=$5.5$, sd=$2.7$, median=$6$]
    \item Did the camera cover make you feel safer about the privacy of your smartphone? (Options: 1 = Not at all safer, 10 = Very much safer) [mean=$5.3$, sd=$2.8$, median=$6$]
    \item Will you continue to use these tools in the future? (Options: 1 = Yes, the privacy filter, 2 = Yes, the camera cover, 3 = Yes, both, 4 = No, both, 5 = Not sure) [removed from the questionnaire and asked during the interview - see question 7 Appendix~\ref{appendix_post_interview}]
    \item What is the main reason for your decision? (Options: 1 = Increased security, 2 = Increased awareness of privacy, 3 = Impractical to use, 4 = Made no real difference, 5 = Other) [removed from the questionnaire and asked during the interview - see question 7 Appendix~\ref{appendix_post_interview}]
\end{enumerate}

\section{Interview Questions}

\subsection{Pre-Intervention Interview Questions}\label{appendix_pre_interview}
\begin{enumerate}
    \item Can you describe a situation where you felt uncomfortable with someone being able to see the screen of your phone? How did you react?
    \item What strategies, if any, do you use to maintain privacy when using your phone in public?
    \item When using your phone in public places, do you tend to choose specific seats (e.g. back against the wall, away from others)? Why or why not?
    \item Are you conscious of who might be looking at your screen when using your phone in a public or social setting?
    \item Have you ever avoided certain activities (eg, reading private messages, checking bank apps) in a social/public setting due to privacy concerns?
    \item How do you feel about people using privacy tools such as screen filters or camera covers?
\end{enumerate}

\subsection{Post-Intervention Interview Questions}\label{appendix_post_interview}
\begin{enumerate}
    \item What was it like for you to use these tools during the period?
    \item Did you notice any changes in how you used your phone in public and/or social settings? If yes, in what way?
    \item Were there situations where you found these tools particularly useful or inconvenient?
    \item How did others react to you using these tools?
    \item Did the tools affect social interactions, for example how often you checked your phone in social situations? If yes, in what way?
    \item To what extent were you self-conscious about using the tools? Did this change over time?
    \item Will you continue to use these tools in the future? Why / why not?
\end{enumerate}

\end{document}